\documentclass[referee]{aa}
\usepackage{amsmath}
\usepackage[dvips]{graphics}
\usepackage {epsfig}
\title{On the variable color of the images of a single source in a gravitational mirage: consequences for the photometric redshift.  }
\author{C. Alard 
        \inst *}

\institute{Institut d'Astrophysique de Paris, 98bis boulevard Arago, 75014
           Paris \\
           \email{alard@iap.fr}}
\date{}
\begin{document}
\abstract {} {In gravitational lensing the average colors of the images are not
identical to the average color of the source. The highly non-linear mapping of
 gravitational lensing does not preserve the color balance of the source, and
 this mapping is different for each image.} 
{The color distortion of the images is illustrated using HST images of the
 lens SL2SJ02140. It is shown that in this lens the color of the images is
 variable, reflecting the variable color of the source.} 
{The average color of the images in SL2SJ02140 are interpreted as a 
variable amplification of different sources regions with different colors.}
{The variation of the average image colors affects the measurements of 
 the photometric redshift of the images. This is especially true for 
SL2SJ02140 where the color variations due to the non-linear mapping of 
the lens simulates pseudo redshifts variations.}
\maketitle
\section{Introduction.}
Gravitational lensing preserve the surface brightness of 
the source the color of the local source element. As a consequence it
is generally assumed that the average color of the images reflects the color
of the source. But this statement is true only if the color of the source
is constant. In practice this is rarely the case as illustrated by SL2SJ02140
(Alard 2009) or SL2S02176 (Alard 2010). Since in a gravitational lens the 
images of the source are 
highly distorted, the surface occupied by an area of the source of a given color
is different in each images. For instance, some source area of a given 
color are over-amplified and thus over represented. This non-linearity
is different in each image, leading to color variations of the images.
\section{SL2SJ02140 as an illustration of the variations in color of the
images of a given source.}
This section will present in more details an analysis of HST data that was
already presented in Alard (2009).
\subsection{Average image colors.}
 The average image colors are computed using all images areas (red+black areas
in Fig. 1). A 3 point Color diagram is computed for each image using the
3 HST photometric bands (see Alard, 2009 for more details). The color
diagrams of the images (Fig. 2) are all different, but the color diagram 
of image C (see Verdugo {\it etal.} 2010 for the image name definitions)
 is quite apart from the color diagrams of images A,B. We will see that
 this difference in color is a consequence of the non-linear mapping
 of the source.
\begin{figure}
\centering{\epsfig{figure=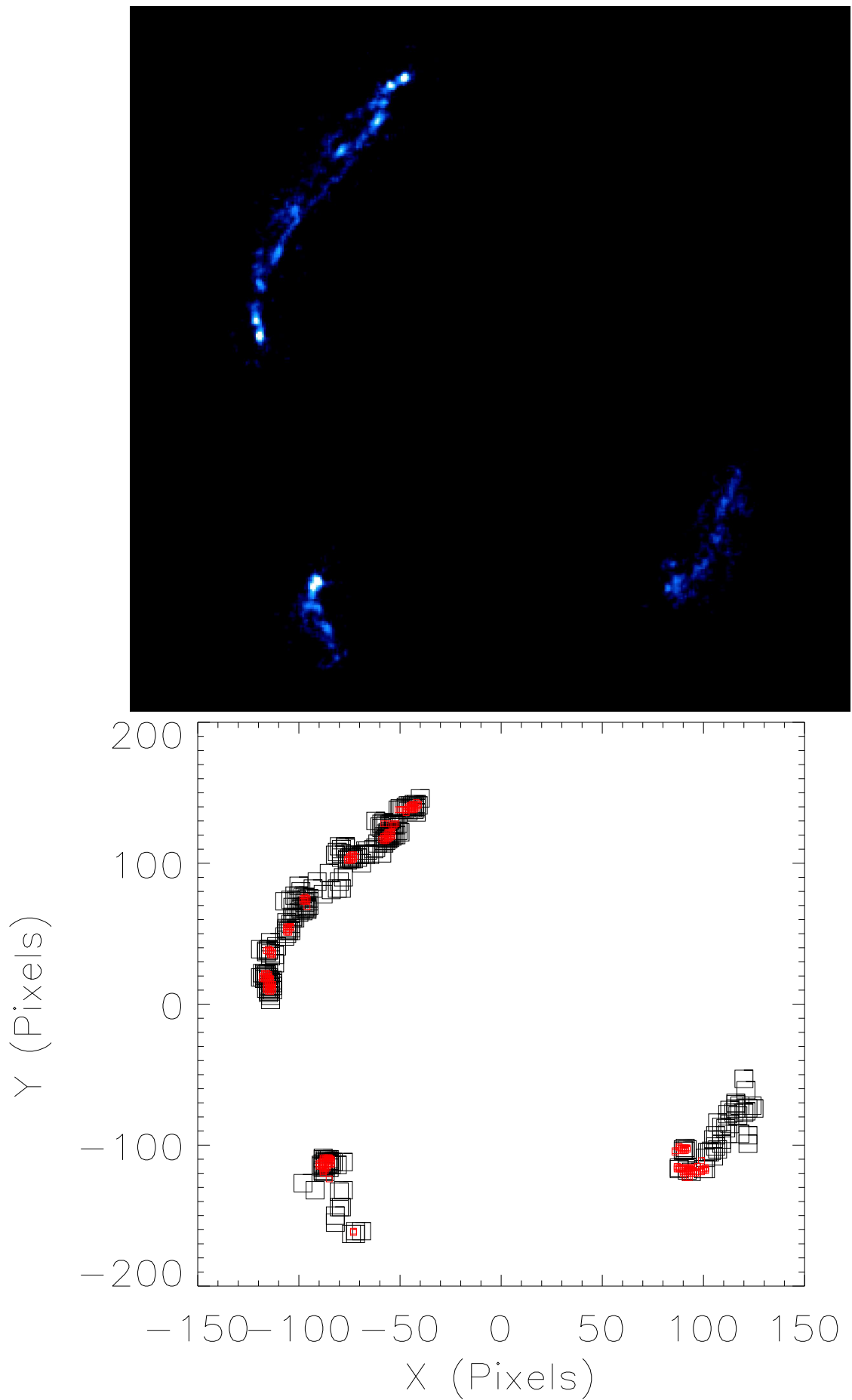,width=16.0cm}}
\caption{The images of the source in SL2SJ02140, upper view
HST image, lower view, image selection diagram. The red areas represent
the bright sharp details (BSD), while the black areas represents fainter
and smoother image areas (SIA).}
\end{figure}
\begin{figure}
\centering{\epsfig{figure=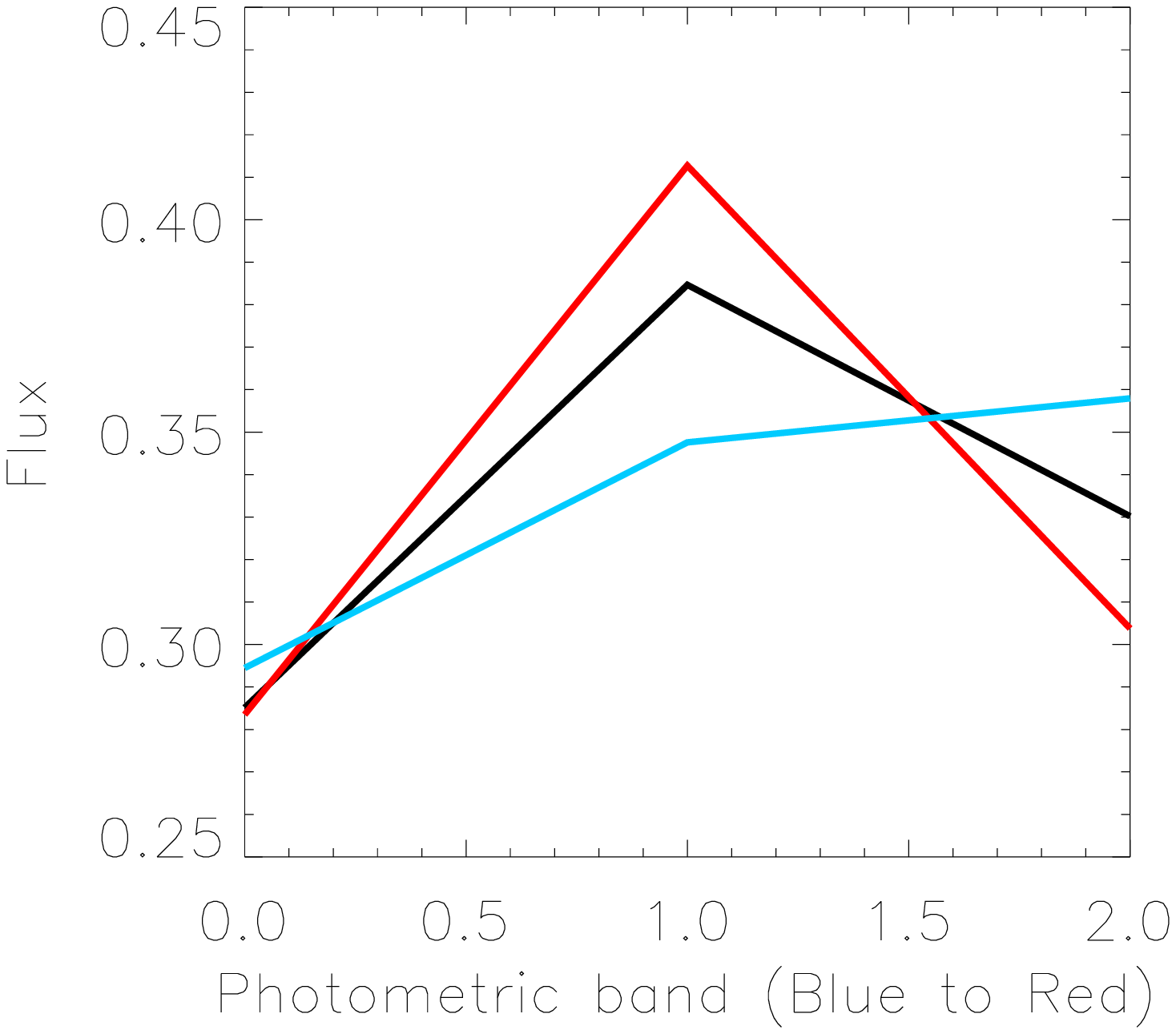,width=10cm}}
\caption{Average image colors. Each point in each diagram represents the
total image flux in a given HST photometric band. Black curve, image A, red image B, and blue image C.}
\end{figure}
\subsection{Detailed analysis of image colors}
 Are the color variations observed in Fig. 2 due to redshift-distance effects
 or are they due to a combination of source color variation and non-linear
 mapping ? This question can be easily answered by analyzing the colors 
variations within each image. If we observe the image of a single source at 
a given redshift, the color of the individual source elements must be similar
in all the images. In SL2SJ02140 the individual image elements are of 2 kinds:
 bright sharp details (BSD) and smoother image areas (SIA). Provided all
the images are from the same source we expect that the BSD and SIA should have all the
same colors in all images. Fig. 3 and Fig. 4 demonstrates that this is just
what we observe. The analysis is facilitated by the fact that the BSD 
and SIA have very different colors and 
the difference in color is very statistically significant (Alard 2009). This result
rules out the possibility that image C is the image of another source. In
such case one should not retrieve similarities in the colors of the BSD
and SIA between image C and A,B. 
\subsection{Interpretation.}
The similarities between the colors of the SIA and BSD demonstrates that
all images comes from the same source. The variations in the image colors
are due to different weightings (amplifications) of the SIA and BSD. Images
A,B are quite dominated by the flux of the BSD, while the weight of the BSD
is much lower in image C. The difference in the weighting of the BSD between the images is just a consequence of the non-linearity of gravitational lens, and
of the complexity of the potential. 
\begin{figure}
\centering{\epsfig{figure=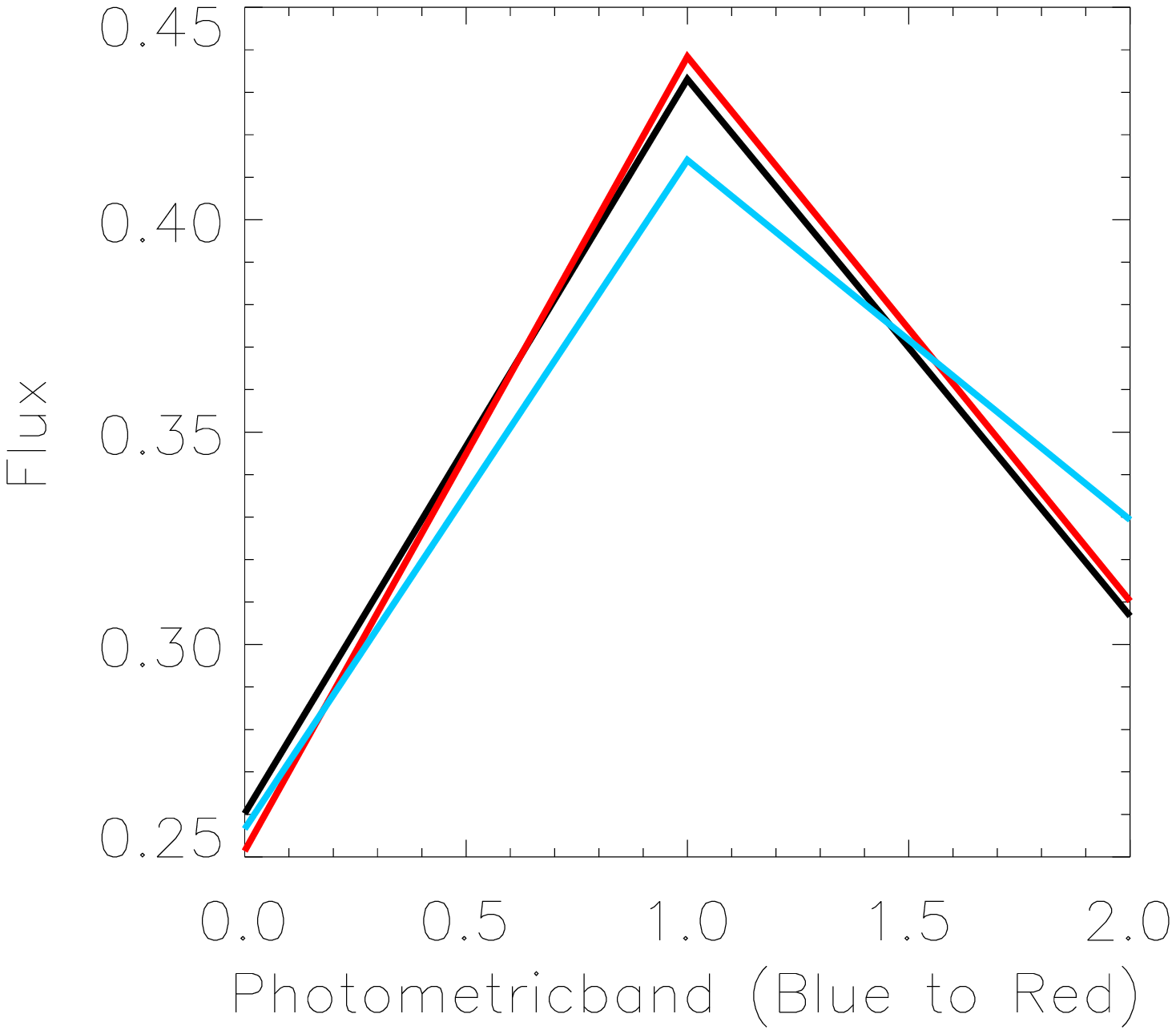,width=10cm}}
\caption{Color of the BSD. Each point in each diagram represents the
total BSD flux in a given HST photometric band. Black curve, image A, red image B, and blue image C. All diagrams are similar within the photometric 
errors.}
\end{figure}
\begin{figure}
\centering{\epsfig{figure=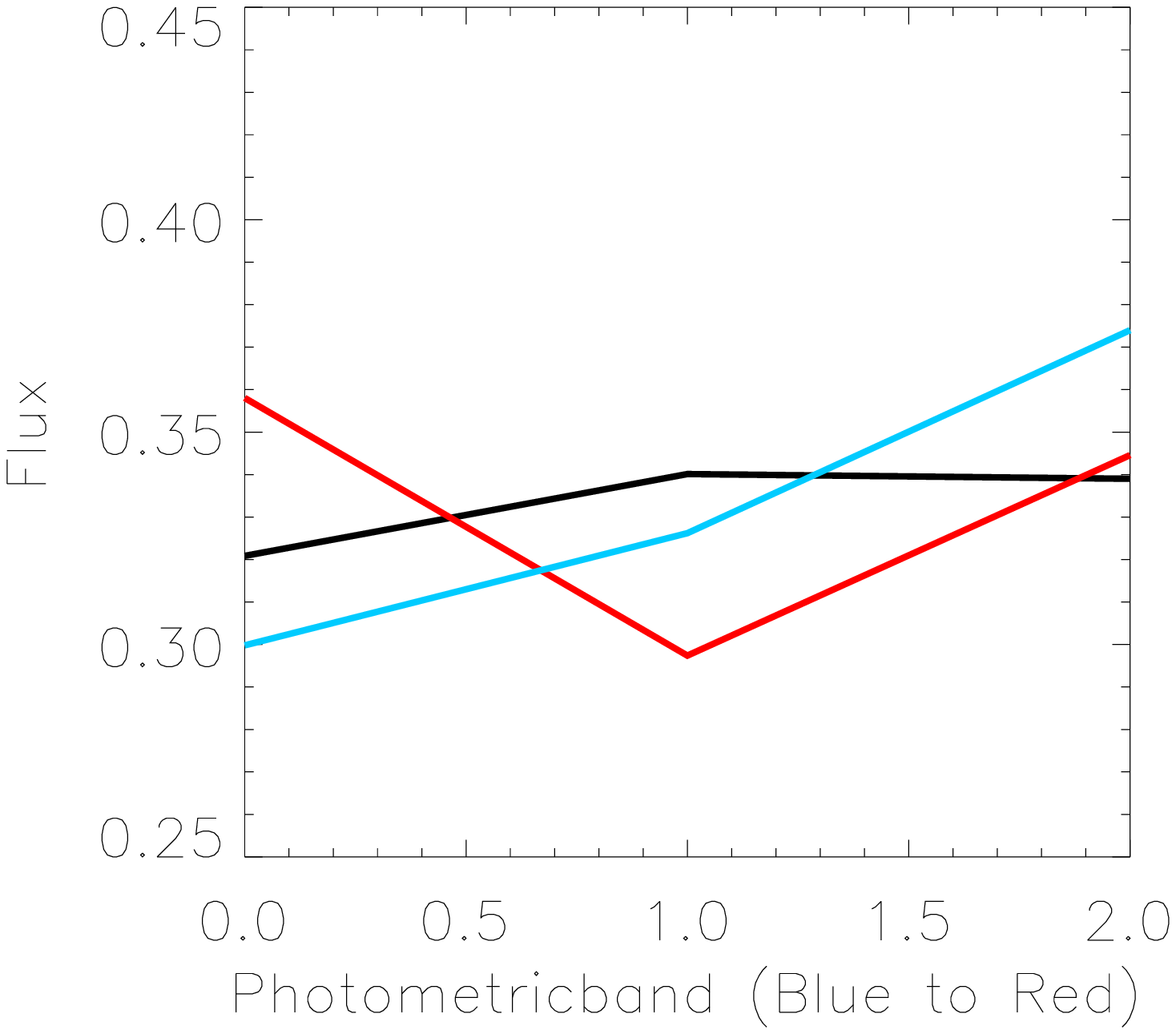,width=10cm}}
\caption{Color of the SIA. Each point in each diagram represents the
total SIA flux in a given HST photometric band. Black curve, image A, red image B, and blue image C. All diagrams are similar within the photometric 
errors.}
\end{figure}
\section{Conclusion}
 The analysis of SL2SJ02140 demonstrates that the
color variations within each image should always be analyzed before
making any conclusion about the average image colors. This analysis is
 absolutely essential before deriving any results about the intrinsic color
of the source. One should not forget that the non-linearity of gravitational
lensing greatly complicates the interpretation of image colors, and that
a straightforward interpretation of the average image colors may be wrong.
For instance Verdugo {\it etal.} 2010 found that using the average color
of the images the pseudo redshift of the source was between 1.0 and 1.8, 
which illustrates the amplitude of the errors on the redshift that can be made when the
non-linearity of the lens mapping is ignored.
\end{document}